\documentclass[aps,prd,twocolumn,groupedaddress,preprintnumbers]{revtex4-2}

\usepackage{amssymb}
\usepackage{amsmath}
\usepackage{graphicx}
\usepackage{enumerate}
\usepackage{mathtools}
\usepackage{color}
\usepackage{bbold}
\usepackage{subfigure}
\usepackage{slashed}
\usepackage[dvipsnames,usenames]{xcolor}
\usepackage{bm}
\usepackage{ulem}
\usepackage{centernot}
\usepackage{bbm}
\usepackage{mathrsfs}

\begin{document}


\preprint{LA-UR-24-21158}

\title{Implications of conservation laws and ergodicity for neutrino flavor instability}



\author{Lucas Johns}
\email[]{ljohns@lanl.gov}
\affiliation{Theoretical Division, Los Alamos National Laboratory, Los Alamos, NM 87545, USA}

\begin{abstract}
Collective neutrino flavor instabilities are believed to be prevalent in core-collapse supernovae and neutron star mergers. This work establishes two points related to instability, both in the spirit of developing a more fundamental understanding of collective flavor dynamics. First, a conservation law related to lepton number implies that spectral crossings are necessary for instability. Second, the most permissive application of the ergodic hypothesis to neutrino flavor evolution implies that spectral crossings are in fact sufficient for instability. We also discuss flavor thermalization and energy conservation in relation to ergodicity.
\end{abstract}

\maketitle

\section{Introduction}

Outcomes of neutrino flavor conversion remain highly uncertain in core-collapse supernovae and neutron star mergers \cite{volpe2024, johns2025neutrino}. A useful approach to the problem is to focus on collective flavor instabilities. In this work we explore the implications of conservation laws and flavor ergodicity for the stability of dense neutrino gases.

Most recent work in this area has concentrated on fast instabilities \cite{sawyer2005}, as distinguished from slow \cite{kostelecky1993neutrino, duan2006collective} and collisional instabilities \cite{johns2021collisional, xiong2023c}. A spectral crossing in a certain distribution related to lepton number is necessary and sufficient for a dense neutrino gas to be unstable to fast flavor conversion \cite{sawyer2016, chakraborty2016c, izaguirre2017, abbar2018, tamborra2021new}. The existing proofs of this statement are based on technical properties of the linearized-system dispersion relation \cite{morinaga2022, dasgupta2022, fiorillo2024theory, dasgupta2025}. Very recently studies have begun to shift the emphasis toward the other instability types \cite{dasgupta2022, fiorillo2024theory, dasgupta2025, fiorillo2025lepton}.

In this paper we prove the necessity of spectral crossings for flavor instability solely by appealing to conservation laws. Linearization is unnecessary. We then give an alternative proof of the sufficiency of spectral crossings for flavor instability using the assumption that neutrino flavor evolution is ergodic \cite{johns2023c, johns2023d, johns2024, johns2025local}. After setting up the general model (Sec.~\ref{sec:gensystem}), we develop the proofs for the fast neutrino system, in which vacuum mixing and collisions are neglected and only fast instabilities are possible (Sec.~\ref{sec:system}). We then adapt the proofs to the general model, in which fast, slow, and collisional instabilities are all potentially present (Sec.~\ref{sec:genan}).

Our purpose in this work is not to prove ergodicity itself, but rather to demonstrate the kinds of implications it would have for flavor instabilities if it were found to be a good approximation for collective neutrino flavor dynamics. In Sec.~\ref{sec:extensions} we consider how the concept of flavor ergodicity could be developed further by linking it to neutrino quantum thermodynamics or applying it to a more restrictive ensemble than is adopted in Secs.~\ref{sec:system} and \ref{sec:genan}. We close the paper with a brief summary of findings in Sec.~\ref{sec:conc}.

\section{The general neutrino system\label{sec:gensystem}}

We focus on the two-flavor scenario and work in terms of polarization vectors $\vec{P}_{\boldsymbol{p}} (t, \boldsymbol{x})$ defined by
\begin{equation}
\rho_{\boldsymbol{p}} (t, \boldsymbol{x}) = \frac{1}{2} \left( P^0_{\boldsymbol{p}} (t, \boldsymbol{x}) \mathbbm{1} + \vec{P}_{\boldsymbol{p}} (t, \boldsymbol{x}) \cdot \vec{\sigma} \right),
\end{equation}
where $\vec{\sigma}$ is a vector of Pauli matrices and $\rho_{\boldsymbol{p}} (t, \boldsymbol{x})$ is the flavor density matrix at time $t$, position $\boldsymbol{x}$, and momentum $\boldsymbol{p}$. We use arrows to distinguish flavor-space vectors from coordinates and momenta. Bars will be used above symbols to denote antineutrino quantities. We will often find it convenient to use the sum and difference vectors
\begin{equation}
\vec{S}_{\boldsymbol{p}} \equiv \vec{P}_{\boldsymbol{p}} + \vec{\bar{P}}_{\boldsymbol{p}}, ~~~ \vec{D}_{\boldsymbol{p}} \equiv \vec{P}_{\boldsymbol{p}} - \vec{\bar{P}}_{\boldsymbol{p}}
\end{equation}
in place of $\vec{P}_{\boldsymbol{p}}$ and $\vec{\bar{P}}_{\boldsymbol{p}}$.

The governing equation of motion for neutrinos is
\begin{align}
    &\left( \partial_t + \boldsymbol{\hat{p}} \cdot \partial_{\boldsymbol{x}} \right) \vec{P}_{\boldsymbol{p}} = \omega_{\boldsymbol{p}} \vec{B} \times \vec{P}_{\boldsymbol{p}} \notag \\
    &\hspace{.7 in} +  \sqrt{2} G_F \big( \vec{D}_0 - \boldsymbol{\hat{p}}  \cdot \vec{\boldsymbol{D}}_1 \big) \times \vec{P}_{\boldsymbol{p}} - \Gamma_{\boldsymbol{p}} \vec{P}_{\boldsymbol{p}}^T. \label{eq:generalP}
\end{align}
In the above, $\omega_{\boldsymbol{p}}$ is the vacuum oscillation frequency, which is defined in terms of the neutrino mass-squared splitting $\delta m^2$ through $\omega_{\boldsymbol{p}} = \delta m^2 / 2 |\boldsymbol{p}|$; $\vec{B}$ is the mass basis vector; $\Gamma_{\boldsymbol{p}}$ is the collisional damping rate; $\vec{P}^T$ is the part of $\vec{P}$ transverse to the flavor axis $\hat{z}$; and the collective difference vectors are
\begin{equation}
\vec{D}_0 \equiv \int \frac{d^3\boldsymbol{q}}{(2\pi)^3} \vec{D}_{\boldsymbol{q}}, ~~~ \vec{\boldsymbol{D}}_1 \equiv \int \frac{d^3\boldsymbol{q}}{(2\pi)^3} \boldsymbol{\hat{q}} \vec{D}_{\boldsymbol{q}}.
\end{equation}
The antineutrino equation of motion is
\begin{align}
    &\left( \partial_t + \boldsymbol{\hat{p}} \cdot \partial_{\boldsymbol{x}} \right) \vec{\bar{P}}_{\boldsymbol{p}} = -\omega_{\boldsymbol{p}} \vec{B} \times \vec{\bar{P}}_{\boldsymbol{p}}\notag \\
    &\hspace{.7 in} +  \sqrt{2} G_F \big( \vec{D}_0 - \boldsymbol{\hat{p}}  \cdot \vec{\boldsymbol{D}}_1 \big) \times \vec{\bar{P}}_{\boldsymbol{p}} - \bar{\Gamma}_{\boldsymbol{p}} \vec{\bar{P}}_{\boldsymbol{p}}^T. \label{eq:generalPbar}
\end{align}
In this study we omit the matter potential and take $\vec{B}$ to be parallel to the flavor axis $\hat{z}$ (the zero-mixing-angle approximation). We treat collisions in a damping approximation, isolating the aspect of collisions that is fundamentally responsible for collisional instability.

In terms of the sum and difference vectors, the equations of motion are
\begin{align}
    \left( \partial_t + \boldsymbol{\hat{p}} \cdot \partial_{\boldsymbol{x}} \right) \vec{S}_{\boldsymbol{p}} = &~\omega_{\boldsymbol{p}} \vec{B} \times \vec{D}_{\boldsymbol{p}} \notag \\
    &+ \sqrt{2} G_F \big( \vec{D}_0 - \boldsymbol{\hat{p}}  \cdot \vec{\boldsymbol{D}}_1 \big) \times \vec{S}_{\boldsymbol{p}} \notag \\
    &- \frac{\Gamma_{\boldsymbol{p}} + \bar{\Gamma}_{\boldsymbol{p}}}{2} \vec{S}_{\boldsymbol{p}}^T - \frac{\Gamma_{\boldsymbol{p}} - \bar{\Gamma}_{\boldsymbol{p}}}{2} \vec{D}_{\boldsymbol{p}}^T \label{eq:generalSp}
\end{align}
and
\begin{align}
    \left( \partial_t + \boldsymbol{\hat{p}} \cdot \partial_{\boldsymbol{x}} \right) \vec{D}_{\boldsymbol{p}} = &~\omega_{\boldsymbol{p}} \vec{B} \times \vec{S}_{\boldsymbol{p}} \notag \\
    &+ \sqrt{2} G_F \big( \vec{D}_0 - \boldsymbol{\hat{p}}  \cdot \vec{\boldsymbol{D}}_1 \big) \times \vec{D}_{\boldsymbol{p}} \notag \\
    &- \frac{\Gamma_{\boldsymbol{p}} + \bar{\Gamma}_{\boldsymbol{p}}}{2} \vec{D}_{\boldsymbol{p}}^T - \frac{\Gamma_{\boldsymbol{p}} - \bar{\Gamma}_{\boldsymbol{p}}}{2} \vec{S}_{\boldsymbol{p}}^T. \label{eq:generalDp}
\end{align}
Let region $\mathcal{R}$ enclose the neutrino system and assume periodic boundary conditions. Define the total difference vector
\begin{equation}
\vec{\mathscr{D}} \equiv \int_\mathcal{R} d^3\boldsymbol{x}~\vec{D}_0.
\end{equation}
Its projection along the flavor axis is an invariant of Eq.~\eqref{eq:generalDp},
\begin{equation}
    \mathscr{D}_z (t) = \textrm{constant}, \label{eq:Dzcons}
\end{equation}
because there is no net flux through the boundary of $\mathcal{R}$ and because $\hat{z}$ is parallel to $\vec{B}$ and perpendicular to $\vec{D}_{\boldsymbol{p}}^T$ and $\vec{S}_{\boldsymbol{p}}^T$.

In the following sections we simplify the general neutrino system [Eqs.~\eqref{eq:generalSp} and \eqref{eq:generalDp}] to the much-studied fast limit, analyze fast instability in relation to conservation laws and ergodicity, and then return to the general system to broaden the analysis to encompass all instability types.

\section{The fast neutrino system\label{sec:system}}

\subsection{Conservation laws and the necessity of an angular crossing}

Fast flavor instability is determined by the $\vec{D}_{\boldsymbol{p}}$ sector, which is uninfluenced by the $\vec{S}_{\boldsymbol{p}}$ dynamics \cite{johns2020}. The relevant equation of motion is therefore
\begin{equation}
\left( \partial_t + \boldsymbol{\hat{p}} \cdot \partial_{\boldsymbol{x}} \right) \vec{D}_{\boldsymbol{p}} = \sqrt{2} G_F \big( \vec{D}_0 - \boldsymbol{\hat{p}}  \cdot \vec{\boldsymbol{D}}_1 \big) \times \vec{D}_{\boldsymbol{p}}. \label{eq:eom0}
\end{equation}
Defining
\begin{equation}
\vec{D}_{\boldsymbol{\hat{p}}} \equiv \int \frac{d|\boldsymbol{p}| |\boldsymbol{p}|^2}{2 \pi^2} \vec{D}_{\boldsymbol{p}},
\end{equation}
we rewrite Eq.~\eqref{eq:eom0} in terms of energy-integrated quantities:
\begin{equation}
\left( \partial_t + \boldsymbol{\hat{p}} \cdot \partial_{\boldsymbol{x}} \right) \vec{D}_{\boldsymbol{\hat{p}}} = \sqrt{2} G_F \big( \vec{D}_0 - \boldsymbol{\hat{p}}  \cdot \vec{\boldsymbol{D}}_1 \big) \times \vec{D}_{\boldsymbol{\hat{p}}} \label{eq:eom}
\end{equation}
with
\begin{equation}
\vec{D}_0 = \int \frac{d\boldsymbol{\hat{q}}}{4 \pi} \vec{D}_{\boldsymbol{\hat{q}}}, ~~ \vec{\boldsymbol{D}}_1 = \int \frac{d\boldsymbol{\hat{q}}}{4 \pi} \boldsymbol{\hat{q}} \vec{D}_{\boldsymbol{\hat{q}}}.
\end{equation}

Under Eq.~\eqref{eq:eom}, difference-vector magnitudes are conserved along particle trajectories:
\begin{equation}
| \vec{D}_{\boldsymbol{\hat{p}}} (t, \boldsymbol{x}_0 + \boldsymbol{\hat{p}} t) | = \textrm{constant}. \label{eq:Dmag}
\end{equation}
The system also obeys the continuity equation
\begin{equation}
\partial_t \vec{D}_0 + \nabla \cdot \vec{\boldsymbol{D}}_1 = 0, \label{eq:cont}
\end{equation}
which follows from integrating Eq.~\eqref{eq:eom} over all propagation angles \cite{zaizen2023}. The integral form is
\begin{equation}
\frac{d\vec{\mathscr{D}}}{dt} + \int_\mathcal{S} d\boldsymbol{\mathcal{S}} \cdot \vec{\boldsymbol{D}}_1 = 0,
\end{equation}
where $\mathcal{S}$ is the surface of $\mathcal{R}$. Then in fact the entirety of $\vec{\mathscr{D}}$ is an invariant, not only its $z$-component as in the general system [Eq.~\eqref{eq:Dzcons}]:
\begin{equation}
\vec{\mathscr{D}} (t) = \textrm{constant}. \label{eq:scrD}
\end{equation}
The total energy of the system is not conserved despite taking the collisionless limit \cite{fiorillo2024}.

Let $\vec{D}_{\boldsymbol{\hat{p}}}^\textrm{i}$ be the flavor state whose stability we are inquiring about. Following precedent, we assume this state to be spatially homogeneous and polarized along the $z$-axis. The system is unstable if it evolves to some state $\vec{D}_{\boldsymbol{\hat{p}}}^\textrm{f} (\boldsymbol{x})$ such that
\begin{equation}
D_{\boldsymbol{\hat{k}},z}^\textrm{f} (\boldsymbol{y}) \neq D_{\boldsymbol{\hat{k}},z}^\textrm{i} \label{eq:D1z}
\end{equation}
for some $\boldsymbol{y}$ and $\boldsymbol{\hat{k}}$. To simplify notation, we define
\begin{equation}
g_{\boldsymbol{\hat{p}}} \equiv D_{\boldsymbol{\hat{p}},z}^\textrm{i}, ~~ \delta g_{\boldsymbol{\hat{p}}} (\boldsymbol{x}) \equiv D_{\boldsymbol{\hat{p}},z}^\textrm{f} (\boldsymbol{x}) - D_{\boldsymbol{\hat{p}},z}^\textrm{i}.
\end{equation}
Equation~\eqref{eq:D1z} becomes
\begin{equation}
\delta g_{\boldsymbol{\hat{k}}} (\boldsymbol{y}) \neq 0. \label{eq:deltag}
\end{equation}
For $\vec{D}_{\boldsymbol{\hat{p}}}^\textrm{f} (\boldsymbol{x})$ to be reachable from $\vec{D}_{\boldsymbol{\hat{p}}}^\textrm{i}$, it must also satisfy
\begin{equation}
    | D_{\boldsymbol{\hat{p}},z}^{\textrm{f}}| \leq | D_{\boldsymbol{\hat{p}},z}^{\textrm{i}}|, \label{eq:DfDi}
\end{equation}
which implies that
\begin{equation}
    \textrm{sgn}[\delta g_{\boldsymbol{\hat{k}}}(\boldsymbol{y})] = - \textrm{sgn}[g_{\boldsymbol{\hat{k}}}]. \label{eq:sgndg}
\end{equation}
The state $\vec{D}_{\boldsymbol{\hat{p}}}^\textrm{i}$ is unstable if and only if Eq.~\eqref{eq:deltag} is satisfied for some $\vec{D}_{\boldsymbol{\hat{p}}}^\textrm{f} (\boldsymbol{x})$ consistent with Eq.~\eqref{eq:sgndg}.

Now we show that an angular crossing (a change in sign of $g_{\boldsymbol{\hat{p}}}$ as a function of $\boldsymbol{\hat{p}}$) is a necessary condition for fast flavor instability. The essential idea is that an angular crossing creates slack in the Eq.~\eqref{eq:scrD} constraint, potentially allowing for flavor conversion.

Suppose that there is no angular crossing. Without loss of generality, take $g_{\boldsymbol{\hat{p}}} \geq 0$ for all $\boldsymbol{\hat{p}}$. Per Eq.~\eqref{eq:scrD}, $\vec{D}_{\boldsymbol{\hat{p}}}^\textrm{f}$ is related to the initial state $\vec{D}_{\boldsymbol{\hat{p}}}^\textrm{i}$ by
\begin{equation}
\mathscr{D}_z^{\textrm{f}} - \mathscr{D}_z^{\textrm{i}} = \int d^3 \boldsymbol{x} \int \frac{d\boldsymbol{\hat{p}}}{4 \pi} \delta g_{\boldsymbol{\hat{p}}} (
\boldsymbol{x}) = 0. \label{eq:deltag0}
\end{equation}
For the initial state to be unstable, there must be some $\boldsymbol{y}$ and $\boldsymbol{\hat{k}}$ such that Eq.~\eqref{eq:deltag} is satisfied. But for Eq.~\eqref{eq:deltag0} to hold, there must also be some $\boldsymbol{y}'$ and $\boldsymbol{\hat{k}}'$ such that
\begin{equation}
\textrm{sgn} [\delta g_{\boldsymbol{\hat{k}}'} (\boldsymbol{y}') ] = - \textrm{sgn} [\delta g_{\boldsymbol{\hat{k}}} (\boldsymbol{y})].
\end{equation}
This implies that either $\delta g_{\boldsymbol{\hat{k}}'} (\boldsymbol{y}') > 0$ or $\delta g_{\boldsymbol{\hat{k}}} (\boldsymbol{y}) > 0$, in defiance of Eq.~\eqref{eq:sgndg}. Therefore fast instability requires an angular crossing.

\subsection{Flavor ergodicity and the sufficiency of an angular crossing\label{sec:fastsuff}}

Now let us assume that flavor evolution is ergodic: Over a sufficiently long duration, the neutrino system visits any flavor state not excluded by conservation laws.

We will be making use of the spatial coarse-graining operator
\begin{equation}
\langle \cdot \rangle \equiv \int_{\mathcal{R}'} \frac{d^3 \boldsymbol{x}'}{V'} (\cdot), \label{eq:cgop}
\end{equation}
where $V'$ is the volume of region $\mathcal{R}'$. Coarse-graining Eq.~\eqref{eq:eom} leads to
\begin{equation}
\left( \partial_t + \boldsymbol{\hat{p}} \cdot \partial_{\boldsymbol{x}} \right) \langle \vec{D}_{\boldsymbol{\hat{p}}} \rangle = \sqrt{2} G_F \big\langle \big( \vec{D}_0 - \boldsymbol{\hat{p}} \cdot \vec{\boldsymbol{D}}_1 \big) \times \vec{D}_{\boldsymbol{\hat{p}}} \big\rangle.
\end{equation}
In general the coarse-grained cross products do not factorize, \textit{e.g.},
\begin{equation}
\langle \vec{\boldsymbol{D}}_1 \times \vec{D}_{\boldsymbol{\hat{p}}} \rangle \neq \langle \vec{\boldsymbol{D}}_1 \rangle \times \langle \vec{D}_{\boldsymbol{\hat{p}}} \rangle.
\end{equation}
As a result, $| \langle \vec{D}_{\boldsymbol{\hat{p}}} \rangle |$ is not conserved even if $\mathcal{R}'$ moves along a neutrino trajectory. This is true for arbitrarily small $V'$. However, by the triangle inequality,
\begin{equation}
| \langle \vec{D}_{\boldsymbol{\hat{p}}} \rangle | \leq \langle | \vec{D}_{\boldsymbol{\hat{p}}} | \rangle. \label{eq:Dtri}
\end{equation}
In contrast, the integral quantity $\vec{\mathscr{D}}$ is unaffected by coarse-graining. Since the conservation laws do not depend on the coarse-graining scale, we choose to coarse-grain over the entire volume: $\mathcal{R}' = \mathcal{R}$. We then have
\begin{equation}
\langle \vec{D}_0 \rangle (t) = \frac{\vec{\mathscr{D}} (t)}{V} = \textrm{constant}. \label{eq:cgD}
\end{equation}

In specifying the ensemble of accessible flavor states, we must decide whether to use the fine-grained or coarse-grained invariants. Fine-grained ergodicity would restrict the ensemble to states sharing the same $| \vec{D}_{\boldsymbol{\hat{p}}} (t, \boldsymbol{x}_0 + \boldsymbol{\hat{p}} t) |$ and $\vec{\mathscr{D}}$. Whether a state is or is not in a particular fine-grained ensemble depends on information at arbitrarily small scales (see the discussion of $| \langle \vec{D}_{\boldsymbol{\hat{p}}} \rangle |$ above). It seems implausible that the overall behavior of a neutrino system should depend on detailed information at extremely minute scales. For this reason, we assume coarse-grained ergodicity. The coarse-grained ensemble is more tractable, consisting of those states $\vec{D}_{\boldsymbol{p}}^{\textrm{f}}$ sharing the same $\langle \vec{D}_0 \rangle$ and satisfying
\begin{equation}
| \langle \vec{D}_{\boldsymbol{\hat{p}}} \rangle | \leq | \langle \vec{D}_{\boldsymbol{\hat{p}}}^\textrm{i} \rangle |, \label{eq:cgDmag}
\end{equation}
where $\vec{D}_{\boldsymbol{\hat{p}}}^\textrm{i} (\boldsymbol{x})$ is a reference flavor state used to define the ensemble. We cannot definitively rule out the possibility that the system admits additional invariants. If this is found to be the case, the ensemble will need to be redefined accordingly. Here we enforce only the constraints specified above, on the basis that no other strictly conserved quantities are known. Further nuances regarding the selection of an ensemble are discussed in Sec.~\ref{sec:altensembles}.

A dynamical answer to the question of stability comes from considering the evolution of states slightly perturbed away from $\vec{D}_{\boldsymbol{\hat{p}}}^\textrm{i}$. The perturbations are allowed to depend on $\boldsymbol{x}$. In linear stability analysis, instability is demonstrated by showing that there exists an exponentially growing solution $\vec{D}_{\boldsymbol{\hat{p}},T} (t, \boldsymbol{x})$ for the part of the vector transverse to the $z$-axis \cite{banerjee2011, chakraborty2016}. In the linearized dynamics, $D_{\boldsymbol{\hat{p}},z}$ is constant with value $D_{\boldsymbol{\hat{p}},z}^\textrm{i}$.

A statistical answer comes from considering the makeup of the ensemble. Assuming flavor ergodicity, instability can be established by showing that the ensemble contains a state $\vec{D}_{\boldsymbol{\hat{p}}}^\textrm{f} (\boldsymbol{x})$ such that Eq.~\eqref{eq:D1z} is satisfied for some $\boldsymbol{\hat{k}}$ and $\boldsymbol{y}$. Equation~\eqref{eq:D1z} ensures that the system eventually evolves to a state not reachable in the linear regime. It serves the same conceptual role as the condition of exponential growth of $\vec{D}_{\boldsymbol{\hat{p}},T}$.

Now we will show that an angular crossing is a sufficient condition for fast instability, using the assumption of ergodic evolution on the coarse-grained ensemble specified above. The proof is conceptually simple because we merely need to establish that nontrivial dynamics is permitted by the constraints. By ergodicity, if some motion is possible, then it is inevitable. The only technical complication is explicitly constructing a suitable $\vec{D}_{\boldsymbol{\hat{p}}}^\textrm{f} (\boldsymbol{x})$.

Suppose that there exist $\boldsymbol{\hat{p}}_+$ and $\boldsymbol{\hat{p}}_-$ such that $g_{\boldsymbol{\hat{p}}_+} > 0$ and $g_{\boldsymbol{\hat{p}}_-} < 0$. Consider the spatially homogeneous state $\vec{D}_{\boldsymbol{\hat{p}}}^\textrm{f}$ that deviates from $\vec{D}_{\boldsymbol{\hat{p}}}^\textrm{i}$ by
\begin{equation}
\delta g_{\boldsymbol{\hat{p}}} = \begin{cases}
- a G_+ (\boldsymbol{\hat{p}}) & \cos \vartheta_+ (\boldsymbol{\hat{p}}) > \cos\Theta \\
+ a G_- (\boldsymbol{\hat{p}}) & \cos \vartheta_- (\boldsymbol{\hat{p}}) > \cos\Theta \\
0 & \textrm{otherwise}
\end{cases}
\end{equation}
with $a > 0$, $\cos\Theta > 0$, $\cos \vartheta_\pm (\boldsymbol{\hat{p}}) \equiv \boldsymbol{\hat{p}} \cdot \boldsymbol{\hat{p}}_\pm$, and
\begin{align}
&G_\pm ( \boldsymbol{\hat{p}} ) \equiv \notag \\
&~~~~\begin{cases}
\exp \left( -\frac{1}{1 - \left( \sin \vartheta_\pm (\boldsymbol{\hat{p}}) / \sin \Theta \right)^2} \right) & \cos \vartheta_\pm (\boldsymbol{\hat{p}}) > \cos\Theta \\
0 & \textrm{otherwise}. \label{eq:Gfxndef}
\end{cases}
\end{align}
Let $a$ and $\Theta$ be small enough that Eq.~\eqref{eq:cgDmag} and the following conditions are satisfied:
\begin{align}
&g_{\boldsymbol{\hat{p}}} > 0 ~~\textrm{for all}~ \boldsymbol{\hat{p}}~ \textrm{such that}~ \cos \vartheta_+ (\boldsymbol{\hat{p}}) > \cos\Theta, \notag \\
&g_{\boldsymbol{\hat{p}}} < 0 ~~\textrm{for all}~ \boldsymbol{\hat{p}}~ \textrm{such that}~ \cos \vartheta_- (\boldsymbol{\hat{p}}) > \cos\Theta.
\end{align}
Then
\begin{align}
\delta g_{0,\pm} &\equiv \mp \int \frac{d\boldsymbol{\hat{p}}}{4\pi} a G_\pm (\boldsymbol{\hat{p}}) \notag\\
&= \mp \frac{a}{2} \int_0^1 du \frac{u \sin^2\Theta}{\sqrt{1-u^2\sin^2\Theta}} \exp \left( - \frac{1}{1-u^2} \right),
\end{align}
from which we see that Eq.~\eqref{eq:cgD} holds: 
\begin{equation}
\mathscr{D}_z^{\textrm{f}} - \mathscr{D}_z^{\textrm{i}} = \int d^3\boldsymbol{x}~(\delta g_{0,+} + \delta g_{0,-}) = 0. 
\end{equation}
Therefore $\vec{D}_{\boldsymbol{\hat{p}}}^\textrm{f}$ is in the ensemble to which $\vec{D}_{\boldsymbol{\hat{p}}}^\textrm{i}$ belongs, and the latter state is unstable.

\section{General analysis of instabilities\label{sec:genan}}

\subsection{Conservation laws and the necessity of a spectral crossing}

In the previous section we analyzed flavor instabilities in the fast limit. We now return to the general neutrino system whose flavor evolution is given by Eqs.~\eqref{eq:generalP} and \eqref{eq:generalPbar} or equivalently by Eqs.~\eqref{eq:generalSp} and \eqref{eq:generalDp}.

Only the projection $\mathscr{D}_z$ is conserved in the general system, rather than the full $\vec{\mathscr{D}}$. This difference is no obstacle to the logic of the preceding section. However, we can no longer isolate the energy-integrated vectors $\vec{D}_{\boldsymbol{\hat{p}}}$  in the analysis. We need to consider both energy-dependent vectors $\vec{P}_{\boldsymbol{p}}$ and $\vec{\bar{P}}_{\boldsymbol{p}}$. For notational convenience, define
\begin{equation}
    f_{E, \boldsymbol{\hat{p}}} \equiv
\begin{cases}
+P_{\boldsymbol{p},z} & E > 0 \\
- \bar{P}_{\boldsymbol{p},z} & E < 0,
\end{cases}
\end{equation}
with $|E| = |\boldsymbol{p}|$. Adapting the notation from before, the state $\lbrace \vec{P}_{\boldsymbol{p}}^{\textrm{i}}, \vec{\bar{P}}_{\boldsymbol{p}}^{\textrm{i}} \rbrace$ is unstable if it evolves to a state $\lbrace \vec{P}_{\boldsymbol{p}}^{\textrm{f}}(\boldsymbol{x}), \vec{\bar{P}}_{\boldsymbol{p}}^{\textrm{f}}(\boldsymbol{x}) \rbrace$ with
\begin{equation}
    P_{\boldsymbol{k},z}^{\textrm{f}}(\boldsymbol{y}) \neq P_{\boldsymbol{k},z}^{\textrm{i}} ~~ \textrm{or} ~~ \bar{P}_{\boldsymbol{k},z}^{\textrm{f}}(\boldsymbol{y}) \neq \bar{P}_{\boldsymbol{k},z}^{\textrm{i}} \label{eq:generalinstPPbar}
\end{equation}
for some $\boldsymbol{y}$ and $\boldsymbol{k}$. Letting
\begin{equation}
    \delta f_{E,\boldsymbol{\hat{p}}}(\boldsymbol{x}) \equiv f_{E,\boldsymbol{\hat{p}}}^{\textrm{f}}(\boldsymbol{x}) - f_{E,\boldsymbol{\hat{p}}}^{\textrm{i}},
\end{equation}
Eq.~\eqref{eq:generalinstPPbar} becomes
\begin{equation}
    \delta f_{\varepsilon,\boldsymbol{\hat{k}}}(\boldsymbol{y}) \neq 0
\end{equation}
with $\varepsilon = \pm | \boldsymbol{k}|$. Because of the constraints
\begin{equation}
    | P_{\boldsymbol{p},z}^{\textrm{f}}| \leq | P_{\boldsymbol{p},z}^{\textrm{i}}|, ~~ | \bar{P}_{\boldsymbol{p},z}^{\textrm{f}}| \leq | \bar{P}_{\boldsymbol{p},z}^{\textrm{i}}|, \label{eq:PPbarmagn}
\end{equation}
the condition
\begin{equation}
    \textrm{sgn} [ \delta f_{\varepsilon,\boldsymbol{\hat{k}}}(\boldsymbol{y})] = - \textrm{sgn} [ f_{\varepsilon,\boldsymbol{\hat{k}}}] \label{eq:sgndf}
\end{equation}
must also be satisfied. 

We will now show that a spectral crossing (a change in sign of $f_{E,\boldsymbol{\hat{p}}}$ as a function of $E$ and $\boldsymbol{\hat{p}}$) is necessary for flavor instability. The proof is structurally the same as in the previous section.

Suppose that there is no spectral crossing. Without loss of generality, take $f_{E,\boldsymbol{\hat{p}}} \geq 0$ for all $E$ and $\boldsymbol{\hat{p}}$. By conservation of $\mathscr{D}_z$, the initial and final states obey
\begin{equation}
    \mathscr{D}_z^{\textrm{f}} - \mathscr{D}_z^{\textrm{i}} = \int d^3 \boldsymbol{x} \int \frac{dE E^2}{2 \pi^2} \int \frac{d\boldsymbol{\hat{p}}}{4\pi} \delta f_{E,\boldsymbol{\hat{p}}} (\boldsymbol{x}) = 0
\end{equation}
with the energy integral running from $E = -\infty$ to $E = +\infty$. But then there must be some $\varepsilon'$, $\boldsymbol{y}'$, and $\boldsymbol{\hat{k}}'$ such that
\begin{equation}
\textrm{sgn} [\delta f_{\varepsilon',\boldsymbol{\hat{k}}'} (\boldsymbol{y}') ] = - \textrm{sgn} [\delta f_{\varepsilon,\boldsymbol{\hat{k}}} (\boldsymbol{y})].
\end{equation}
This implies that either $\delta f_{\boldsymbol{\hat{k}}'} (\boldsymbol{y}') > 0$ or $\delta f_{\boldsymbol{\hat{k}}} (\boldsymbol{y}) > 0$, in defiance of Eq.~\eqref{eq:sgndf}. Therefore flavor instability requires a spectral crossing.

\subsection{Flavor ergodicity and the sufficiency of a spectral crossing\label{sec:gensuff}}

Here we adapt the sufficiency proof in the fast limit to apply to the general system. Suppose that there exist $(E_+, \boldsymbol{\hat{p}}_+)$ and $(E_-, \boldsymbol{\hat{p}}_-)$ such that $f_{E_+,\boldsymbol{\hat{p}}_+}>0$ and $f_{E_-,\boldsymbol{\hat{p}}_-}<0$. Consider the spatially homogeneous state $f_{E,\boldsymbol{\hat{p}}}^{\textrm{f}}$ specified by
\begin{align}
&\delta f_{E,\boldsymbol{\hat{p}}} = \notag \\
&~~~~~~\begin{cases}
- a F_+(E) G_+ (\boldsymbol{\hat{p}}) & \cos \vartheta_+ (\boldsymbol{\hat{p}}) > \cos\Theta\\
& \textrm{and}~|E-E_+| < \Delta E \\
+ a F_-(E) G_- (\boldsymbol{\hat{p}}) & \cos \vartheta_- (\boldsymbol{\hat{p}}) > \cos\Theta\\
& \textrm{and}~|E-E_-| < \Delta E \\
0 & \textrm{otherwise}
\end{cases}
\end{align}
with $a$, $\cos\Theta$, and $\Delta E$ all positive real numbers and
\begin{align}
&F_\pm ( E ) \equiv \notag \\
&~~~~\begin{cases}
\frac{1}{E^2} \exp \left( -\frac{1}{1 - \left( (E- E_{\pm})/\Delta E \right)^2} \right) & \frac{|E - E_{\pm}|}{\Delta E} < 1 \\
0 & \textrm{otherwise}.
\end{cases}
\end{align}
The function $G_{\pm}$ was defined in Eq.~\eqref{eq:Gfxndef}. Let $a$, $\Theta$, and $\Delta E$ be small enough that Eq.~\eqref{eq:PPbarmagn} and the following conditions are satisfied:
\begin{align}
&f_{E,\boldsymbol{\hat{p}}} > 0 ~~\textrm{for all}~ E, \boldsymbol{\hat{p}}~ \textrm{such that}~ \cos \vartheta_+ (\boldsymbol{\hat{p}}) > \cos\Theta \notag\\
&\hspace{1.5 in}\textrm{and}~ |E - E_+| < \Delta E, \notag \\
&f_{E,\boldsymbol{\hat{p}}} < 0 ~~\textrm{for all}~ E, \boldsymbol{\hat{p}}~ \textrm{such that}~ \cos \vartheta_- (\boldsymbol{\hat{p}}) > \cos\Theta, \notag\\
&\hspace{1.5 in}\textrm{and}~ |E - E_-| < \Delta E. \notag \\
\end{align}
Then
\begin{align}
\delta g_{0,\pm} &\equiv \mp \int \frac{dE E^2}{2 \pi^2} \int \frac{d\boldsymbol{\hat{p}}}{4\pi} a F_{\pm}(E) G_\pm (\boldsymbol{\hat{p}}) \notag\\
&= \mp \frac{a \Delta E}{4 \pi^2} \int_{-1}^{1} dv \exp \left( - \frac{1}{1-v^2} \right) \notag\\
&\hspace{.17 in}\times\int_0^1 du \frac{u \sin^2\Theta}{\sqrt{1-u^2\sin^2\Theta}} \exp \left( - \frac{1}{1-u^2} \right),
\end{align}
and so
\begin{equation}
\mathscr{D}_z^{\textrm{f}} - \mathscr{D}_z^{\textrm{i}} = \int d^3\boldsymbol{x}~(\delta g_{0,+} + \delta g_{0,-}) = 0. 
\end{equation}
Therefore the state with polarization vectors $\lbrace \vec{P}_{\boldsymbol{\hat{p}}}^\textrm{f}, \vec{\bar{P}}_{\boldsymbol{\hat{p}}}^\textrm{f} \rbrace$ is in the ensemble to which the state with $\lbrace \vec{P}_{\boldsymbol{\hat{p}}}^\textrm{i}, \vec{\bar{P}}_{\boldsymbol{\hat{p}}}^\textrm{i} \rbrace$ belongs. By ergodicity, we conclude that the latter state is unstable.

\section{Possible extensions\label{sec:extensions}}

\subsection{Instability as nonmaximal entropy}

In applying ergodicity, our approach has been to identify a state differing in its polarization $z$-components from a given initial state, yet still sharing the same value of $\mathscr{D}_z$. In linear stability analysis, the $z$-components are constant. They change only if the transverse polarization components grow out of the linear regime. These two types of analysis are complementary---though of course the first is only valuable to the extent that ergodicity is in fact a valid approximation.

The ergodic analysis is aligned with a third viewpoint on flavor instability, which posits that fixed points are unstable if they are not at maximal coarse-grained entropy \cite{johns2023c}. The coarse-grained entropy density associated with $\rho_{\boldsymbol{p}}$ is
\begin{equation}
    s_{\boldsymbol{p}} = -\textrm{Tr} \left[ \langle \rho_{\boldsymbol{p}} \rangle \log \langle \rho_{\boldsymbol{p}} \rangle + (1 - \langle \rho_{\boldsymbol{p}} \rangle) \log (1 - \langle \rho_{\boldsymbol{p}} \rangle) \right].
\end{equation}
A decrease in $| \langle \vec{P}_{\boldsymbol{p}} \rangle|$ coincides with an increase in $s_{\boldsymbol{p}}$ \cite{johns2025local}. The standard initial conditions in stability analysis have all $\vec{P}_{\boldsymbol{p}}$ spatially homogeneous and aligned with $\pm \hat{z}$, apart from perturbations. Taking the perturbations to be vanishingly small and ignoring collisions, each polarization vector maintains
\begin{equation}
    | \langle \vec{P}_{\boldsymbol{p}} \rangle| \leq \langle |\vec{P}_{\boldsymbol{p}}|\rangle
\end{equation}
throughout its evolution, implying
\begin{equation}
    s_{\boldsymbol{p}}(t) \geq s_{\boldsymbol{p}}(t_0)
\end{equation}
for all $t \geq t_0$. Since the transverse parts initially vanish by assumption, the only way for entropy to increase is for $|P_{\boldsymbol{p},z}|$ to decrease. In the preceding section this was precisely the question we asked: Is it possible to decrease $|P_{\boldsymbol{p},z}|$ without changing $\mathscr{D}_z$? The ergodic analysis introduced in this paper is closely related to asking whether entropy can increase subject to constraints.

We could base the analysis of flavor instability not on ergodicity per se but rather on flavor thermalization, the hypothesis that neutrinos maximize coarse-grained entropy. Like ergodicity, flavor thermalization may or may not be realized in dense neutrino systems. The question of which conserved quantities should be adopted arises in the entropic approach as well.

\subsection{More restrictive ensembles\label{sec:altensembles}}

Throughout this work we have assumed that the ergodic manifold consists of states with only $\mathscr{D}_z$ or $\vec{\mathscr{D}}$ as an invariant. But as Lichtenberg and Lieberman write in their classic textbook, ``In a sense, ergodicity is universal, and the central question is to define the subspace over which it exists'' \cite{lichtenberg1992regular}.

In Sec.~\ref{sec:fastsuff} we argued that the fine-grained invariants $|\vec{D}_{\boldsymbol{p}}(t, \boldsymbol{x}_0 + \boldsymbol{\hat{p}} t) |$ should be overlooked on the grounds that the dependence of the coarse-grained behavior of the neutrino system on the particular fine-grained values is implausible. This is not to say that the existence of infinitely many fine-grained invariants cannot influence the overall dynamics. But if there is an effect, it is more likely that the fine-grained invariants impede the useful application of ergodicity altogether than suggest a more restrictive ensemble accommodating these invariants.

What we have not yet discussed is the possibility that ergodicity should be assumed to hold only on a manifold of constant energy. On the one hand, energy is not strictly conserved in neutrino quantum kinetics even in the collisionless limit \cite{fiorillo2024}. But on the other hand, it may be approximately conserved at stable equilibria \cite{johns2023c, johns2024}.

A recent theoretical approach to coarse-grained neutrino flavor dynamics treats instabilities as collision-like interactions between the mean $\langle \rho\rangle$ and subgrid (Fourier mode $\boldsymbol{k} \neq 0$) fluctuations in $\rho$, the latter being regarded as quasiparticle degrees of freedom called \textit{flavor waves}  \cite{fiorillo2025collective, johns2025local}. This approach motivates consideration of the energies $U_{\textrm{cg}}$ and $U_{\textrm{fw}}$ in $\langle \rho\rangle$ and the flavor waves, respectively. An analysis like the one in this paper could be developed with the ergodic manifold taken to be at fixed $U_{\textrm{cg}} + U_{\textrm{fw}}$ in addition to fixed $\mathscr{D}_z$ or $\vec{\mathscr{D}}$. Recent studies show that spectral crossings are not sufficient for instability, in contradiction with Sec.~\ref{sec:gensuff} \cite{dasgupta2025, fiorillo2025lepton}. Future work is required to determine whether this finding of insufficiency is compatible with ergodicity in a more restrictive ensemble.

\section{Summary\label{sec:conc}}

We have explored collective flavor instabilities in relation to two fundamental concepts---conservation laws and ergodicity---and made two main points. One, conservation of $\mathscr{D}_z$ prohibits instability unless there is a spectral crossing. Two, if flavor conversion is ergodic on a manifold of fixed $\mathscr{D}_z$ and no other invariants, then a spectral crossing is sufficient for instability.

As discussed in Sec.~\ref{sec:extensions}, the fixed-$\mathscr{D}_z$ ensemble is only the most permissive selection one might make. Ergodicity might obtain only on a more restrictive set of states consistent with additional invariants, and the sufficiency proof presented above would need to be reevaluated accordingly. Or, of course, flavor evolution may be nonergodic altogether. If this is the case, it will be important to understand the nature of the obstruction.

\begin{acknowledgments}
This work was supported by a Feynman Fellowship through LANL LDRD project number 20230788PRD1.
\end{acknowledgments}

\bibliography{all_papers}

\begin{thebibliography}{28}%
\makeatletter
\providecommand \@ifxundefined [1]{%
 \@ifx{#1\undefined}
}%
\providecommand \@ifnum [1]{%
 \ifnum #1\expandafter \@firstoftwo
 \else \expandafter \@secondoftwo
 \fi
}%
\providecommand \@ifx [1]{%
 \ifx #1\expandafter \@firstoftwo
 \else \expandafter \@secondoftwo
 \fi
}%
\providecommand \natexlab [1]{#1}%
\providecommand \enquote  [1]{``#1''}%
\providecommand \bibnamefont  [1]{#1}%
\providecommand \bibfnamefont [1]{#1}%
\providecommand \citenamefont [1]{#1}%
\providecommand \href@noop [0]{\@secondoftwo}%
\providecommand \href [0]{\begingroup \@sanitize@url \@href}%
\providecommand \@href[1]{\@@startlink{#1}\@@href}%
\providecommand \@@href[1]{\endgroup#1\@@endlink}%
\providecommand \@sanitize@url [0]{\catcode `\\12\catcode `\$12\catcode `\&12\catcode `\#12\catcode `\^12\catcode `\_12\catcode `\%12\relax}%
\providecommand \@@startlink[1]{}%
\providecommand \@@endlink[0]{}%
\providecommand \url  [0]{\begingroup\@sanitize@url \@url }%
\providecommand \@url [1]{\endgroup\@href {#1}{\urlprefix }}%
\providecommand \urlprefix  [0]{URL }%
\providecommand \Eprint [0]{\href }%
\providecommand \doibase [0]{https://doi.org/}%
\providecommand \selectlanguage [0]{\@gobble}%
\providecommand \bibinfo  [0]{\@secondoftwo}%
\providecommand \bibfield  [0]{\@secondoftwo}%
\providecommand \translation [1]{[#1]}%
\providecommand \BibitemOpen [0]{}%
\providecommand \bibitemStop [0]{}%
\providecommand \bibitemNoStop [0]{.\EOS\space}%
\providecommand \EOS [0]{\spacefactor3000\relax}%
\providecommand \BibitemShut  [1]{\csname bibitem#1\endcsname}%
\let\auto@bib@innerbib\@empty
\bibitem [{\citenamefont {Volpe}(2024)}]{volpe2024}%
  \BibitemOpen
  \bibfield  {author} {\bibinfo {author} {\bibfnamefont {M.~C.}\ \bibnamefont {Volpe}},\ }\bibfield  {title} {\bibinfo {title} {Neutrinos from dense environments: Flavor mechanisms, theoretical approaches, observations, and new directions},\ }\href {https://doi.org/10.1103/RevModPhys.96.025004} {\bibfield  {journal} {\bibinfo  {journal} {Rev. Mod. Phys.}\ }\textbf {\bibinfo {volume} {96}},\ \bibinfo {pages} {025004} (\bibinfo {year} {2024})}\BibitemShut {NoStop}%
\bibitem [{\citenamefont {Johns}\ \emph {et~al.}(2025)\citenamefont {Johns}, \citenamefont {Richers},\ and\ \citenamefont {Wu}}]{johns2025neutrino}%
  \BibitemOpen
  \bibfield  {author} {\bibinfo {author} {\bibfnamefont {L.}~\bibnamefont {Johns}}, \bibinfo {author} {\bibfnamefont {S.}~\bibnamefont {Richers}},\ and\ \bibinfo {author} {\bibfnamefont {M.-R.}\ \bibnamefont {Wu}},\ }\bibfield  {title} {\bibinfo {title} {Neutrino oscillations in core-collapse supernovae and neutron star mergers},\ }\href {https://www.annualreviews.org/content/journals/10.1146/annurev-nucl-121423-100853} {\bibfield  {journal} {\bibinfo  {journal} {Annu. Rev. Nucl. Part. Sci.}\ } (\bibinfo {year} {2025})}\BibitemShut {NoStop}%
\bibitem [{\citenamefont {Sawyer}(2005)}]{sawyer2005}%
  \BibitemOpen
  \bibfield  {author} {\bibinfo {author} {\bibfnamefont {R.~F.}\ \bibnamefont {Sawyer}},\ }\bibfield  {title} {\bibinfo {title} {Speed-up of neutrino transformations in a supernova environment},\ }\href {https://doi.org/10.1103/PhysRevD.72.045003} {\bibfield  {journal} {\bibinfo  {journal} {Phys. Rev. D}\ }\textbf {\bibinfo {volume} {72}},\ \bibinfo {pages} {045003} (\bibinfo {year} {2005})}\BibitemShut {NoStop}%
\bibitem [{\citenamefont {Kosteleck{\`y}}\ and\ \citenamefont {Samuel}(1993)}]{kostelecky1993neutrino}%
  \BibitemOpen
  \bibfield  {author} {\bibinfo {author} {\bibfnamefont {V.~A.}\ \bibnamefont {Kosteleck{\`y}}}\ and\ \bibinfo {author} {\bibfnamefont {S.}~\bibnamefont {Samuel}},\ }\bibfield  {title} {\bibinfo {title} {Neutrino oscillations in the early universe with an inverted neutrino-mass hierarchy},\ }\href@noop {} {\bibfield  {journal} {\bibinfo  {journal} {Phys. Lett. B}\ }\textbf {\bibinfo {volume} {318}},\ \bibinfo {pages} {127} (\bibinfo {year} {1993})}\BibitemShut {NoStop}%
\bibitem [{\citenamefont {Duan}\ \emph {et~al.}(2006)\citenamefont {Duan}, \citenamefont {Fuller},\ and\ \citenamefont {Qian}}]{duan2006collective}%
  \BibitemOpen
  \bibfield  {author} {\bibinfo {author} {\bibfnamefont {H.}~\bibnamefont {Duan}}, \bibinfo {author} {\bibfnamefont {G.~M.}\ \bibnamefont {Fuller}},\ and\ \bibinfo {author} {\bibfnamefont {Y.-Z.}\ \bibnamefont {Qian}},\ }\bibfield  {title} {\bibinfo {title} {Collective neutrino flavor transformation in supernovae},\ }\href@noop {} {\bibfield  {journal} {\bibinfo  {journal} {Phys. Rev. D}\ }\textbf {\bibinfo {volume} {74}},\ \bibinfo {pages} {123004} (\bibinfo {year} {2006})}\BibitemShut {NoStop}%
\bibitem [{\citenamefont {Johns}(2023{\natexlab{a}})}]{johns2021collisional}%
  \BibitemOpen
  \bibfield  {author} {\bibinfo {author} {\bibfnamefont {L.}~\bibnamefont {Johns}},\ }\bibfield  {title} {\bibinfo {title} {{Collisional Flavor Instabilities of Supernova Neutrinos}},\ }\href {https://doi.org/10.1103/PhysRevLett.130.191001} {\bibfield  {journal} {\bibinfo  {journal} {Phys. Rev. Lett.}\ }\textbf {\bibinfo {volume} {130}},\ \bibinfo {pages} {191001} (\bibinfo {year} {2023}{\natexlab{a}})},\ \Eprint {https://arxiv.org/abs/2104.11369} {arXiv:2104.11369 [hep-ph]} \BibitemShut {NoStop}%
\bibitem [{\citenamefont {Xiong}\ \emph {et~al.}(2023)\citenamefont {Xiong}, \citenamefont {Johns}, \citenamefont {Wu},\ and\ \citenamefont {Duan}}]{xiong2023c}%
  \BibitemOpen
  \bibfield  {author} {\bibinfo {author} {\bibfnamefont {Z.}~\bibnamefont {Xiong}}, \bibinfo {author} {\bibfnamefont {L.}~\bibnamefont {Johns}}, \bibinfo {author} {\bibfnamefont {M.-R.}\ \bibnamefont {Wu}},\ and\ \bibinfo {author} {\bibfnamefont {H.}~\bibnamefont {Duan}},\ }\bibfield  {title} {\bibinfo {title} {Collisional flavor instability in dense neutrino gases},\ }\href {https://doi.org/10.1103/PhysRevD.108.083002} {\bibfield  {journal} {\bibinfo  {journal} {Phys. Rev. D}\ }\textbf {\bibinfo {volume} {108}},\ \bibinfo {pages} {083002} (\bibinfo {year} {2023})}\BibitemShut {NoStop}%
\bibitem [{\citenamefont {Sawyer}(2016)}]{sawyer2016}%
  \BibitemOpen
  \bibfield  {author} {\bibinfo {author} {\bibfnamefont {R.~F.}\ \bibnamefont {Sawyer}},\ }\bibfield  {title} {\bibinfo {title} {Neutrino cloud instabilities just above the neutrino sphere of a supernova},\ }\href {https://doi.org/10.1103/PhysRevLett.116.081101} {\bibfield  {journal} {\bibinfo  {journal} {Phys. Rev. Lett.}\ }\textbf {\bibinfo {volume} {116}},\ \bibinfo {pages} {081101} (\bibinfo {year} {2016})}\BibitemShut {NoStop}%
\bibitem [{\citenamefont {Chakraborty}\ \emph {et~al.}(2016{\natexlab{a}})\citenamefont {Chakraborty}, \citenamefont {Hansen}, \citenamefont {Izaguirre},\ and\ \citenamefont {Raffelt}}]{chakraborty2016c}%
  \BibitemOpen
  \bibfield  {author} {\bibinfo {author} {\bibfnamefont {S.}~\bibnamefont {Chakraborty}}, \bibinfo {author} {\bibfnamefont {R.~S.}\ \bibnamefont {Hansen}}, \bibinfo {author} {\bibfnamefont {I.}~\bibnamefont {Izaguirre}},\ and\ \bibinfo {author} {\bibfnamefont {G.}~\bibnamefont {Raffelt}},\ }\bibfield  {title} {\bibinfo {title} {Self-induced neutrino flavor conversion without flavor mixing},\ }\href {https://doi.org/10.1088/1475-7516/2016/03/042} {\bibfield  {journal} {\bibinfo  {journal} {J. Cosmol. Astropart. Phys.}\ }\textbf {\bibinfo {volume} {2016}}\bibinfo  {number} { (03)},\ \bibinfo {pages} {042}}\BibitemShut {NoStop}%
\bibitem [{\citenamefont {Izaguirre}\ \emph {et~al.}(2017)\citenamefont {Izaguirre}, \citenamefont {Raffelt},\ and\ \citenamefont {Tamborra}}]{izaguirre2017}%
  \BibitemOpen
\bibfield  {number} {  }\bibfield  {author} {\bibinfo {author} {\bibfnamefont {I.}~\bibnamefont {Izaguirre}}, \bibinfo {author} {\bibfnamefont {G.}~\bibnamefont {Raffelt}},\ and\ \bibinfo {author} {\bibfnamefont {I.}~\bibnamefont {Tamborra}},\ }\bibfield  {title} {\bibinfo {title} {Fast pairwise conversion of supernova neutrinos: A dispersion relation approach},\ }\href {https://doi.org/10.1103/PhysRevLett.118.021101} {\bibfield  {journal} {\bibinfo  {journal} {Phys. Rev. Lett.}\ }\textbf {\bibinfo {volume} {118}},\ \bibinfo {pages} {021101} (\bibinfo {year} {2017})}\BibitemShut {NoStop}%
\bibitem [{\citenamefont {Abbar}\ and\ \citenamefont {Duan}(2018)}]{abbar2018}%
  \BibitemOpen
  \bibfield  {author} {\bibinfo {author} {\bibfnamefont {S.}~\bibnamefont {Abbar}}\ and\ \bibinfo {author} {\bibfnamefont {H.}~\bibnamefont {Duan}},\ }\bibfield  {title} {\bibinfo {title} {Fast neutrino flavor conversion: Roles of dense matter and spectrum crossing},\ }\href {https://doi.org/10.1103/PhysRevD.98.043014} {\bibfield  {journal} {\bibinfo  {journal} {Phys. Rev. D}\ }\textbf {\bibinfo {volume} {98}},\ \bibinfo {pages} {043014} (\bibinfo {year} {2018})}\BibitemShut {NoStop}%
\bibitem [{\citenamefont {Tamborra}\ and\ \citenamefont {Shalgar}(2021)}]{tamborra2021new}%
  \BibitemOpen
  \bibfield  {author} {\bibinfo {author} {\bibfnamefont {I.}~\bibnamefont {Tamborra}}\ and\ \bibinfo {author} {\bibfnamefont {S.}~\bibnamefont {Shalgar}},\ }\bibfield  {title} {\bibinfo {title} {New developments in flavor evolution of a dense neutrino gas},\ }\href@noop {} {\bibfield  {journal} {\bibinfo  {journal} {Annu. Rev. Nucl. Part. Sci.}\ }\textbf {\bibinfo {volume} {71}},\ \bibinfo {pages} {165} (\bibinfo {year} {2021})}\BibitemShut {NoStop}%
\bibitem [{\citenamefont {Morinaga}(2022)}]{morinaga2022}%
  \BibitemOpen
  \bibfield  {author} {\bibinfo {author} {\bibfnamefont {T.}~\bibnamefont {Morinaga}},\ }\bibfield  {title} {\bibinfo {title} {Fast neutrino flavor instability and neutrino flavor lepton number crossings},\ }\href {https://doi.org/10.1103/PhysRevD.105.L101301} {\bibfield  {journal} {\bibinfo  {journal} {Phys. Rev. D}\ }\textbf {\bibinfo {volume} {105}},\ \bibinfo {pages} {L101301} (\bibinfo {year} {2022})}\BibitemShut {NoStop}%
\bibitem [{\citenamefont {Dasgupta}(2022)}]{dasgupta2022}%
  \BibitemOpen
  \bibfield  {author} {\bibinfo {author} {\bibfnamefont {B.}~\bibnamefont {Dasgupta}},\ }\bibfield  {title} {\bibinfo {title} {Collective neutrino flavor instability requires a crossing},\ }\href {https://doi.org/10.1103/PhysRevLett.128.081102} {\bibfield  {journal} {\bibinfo  {journal} {Phys. Rev. Lett.}\ }\textbf {\bibinfo {volume} {128}},\ \bibinfo {pages} {081102} (\bibinfo {year} {2022})}\BibitemShut {NoStop}%
\bibitem [{\citenamefont {Fiorillo}\ and\ \citenamefont {Raffelt}(2024)}]{fiorillo2024theory}%
  \BibitemOpen
  \bibfield  {author} {\bibinfo {author} {\bibfnamefont {D.~F.~G.}\ \bibnamefont {Fiorillo}}\ and\ \bibinfo {author} {\bibfnamefont {G.~G.}\ \bibnamefont {Raffelt}},\ }\bibfield  {title} {\bibinfo {title} {Theory of neutrino fast flavor evolution. i. linear response theory and stability conditions},\ }\href@noop {} {\bibfield  {journal} {\bibinfo  {journal} {J. High Energy Phys.}\ }\textbf {\bibinfo {volume} {2024}}\bibinfo  {number} { (225)}}\BibitemShut {NoStop}%
\bibitem [{\citenamefont {Dasgupta}\ and\ \citenamefont {Mukherjee}(2025)}]{dasgupta2025}%
  \BibitemOpen
\bibfield  {number} {  }\bibfield  {author} {\bibinfo {author} {\bibfnamefont {B.}~\bibnamefont {Dasgupta}}\ and\ \bibinfo {author} {\bibfnamefont {D.}~\bibnamefont {Mukherjee}},\ }\bibfield  {title} {\bibinfo {title} {Sufficient and necessary conditions for collective neutrino instability: Fast, slow, and mixed},\ }\href@noop {} {\bibfield  {journal} {\bibinfo  {journal} {arXiv preprint arXiv:2505.03886}\ } (\bibinfo {year} {2025})}\BibitemShut {NoStop}%
\bibitem [{\citenamefont {Fiorillo}\ and\ \citenamefont {Raffelt}(2025{\natexlab{a}})}]{fiorillo2025lepton}%
  \BibitemOpen
  \bibfield  {author} {\bibinfo {author} {\bibfnamefont {D.~F.}\ \bibnamefont {Fiorillo}}\ and\ \bibinfo {author} {\bibfnamefont {G.~G.}\ \bibnamefont {Raffelt}},\ }\bibfield  {title} {\bibinfo {title} {Lepton number crossings are insufficient for flavor instabilities},\ }\href@noop {} {\bibfield  {journal} {\bibinfo  {journal} {arXiv preprint arXiv:2507.22987}\ } (\bibinfo {year} {2025}{\natexlab{a}})}\BibitemShut {NoStop}%
\bibitem [{\citenamefont {Johns}(2023{\natexlab{b}})}]{johns2023c}%
  \BibitemOpen
  \bibfield  {author} {\bibinfo {author} {\bibfnamefont {L.}~\bibnamefont {Johns}},\ }\bibfield  {title} {\bibinfo {title} {Thermodynamics of oscillating neutrinos},\ }\href@noop {} {\  (\bibinfo {year} {2023}{\natexlab{b}})},\ \Eprint {https://arxiv.org/abs/2306.14982} {arXiv:2306.14982 [hep-ph]} \BibitemShut {NoStop}%
\bibitem [{\citenamefont {Johns}\ and\ \citenamefont {Rodriguez}(2023)}]{johns2023d}%
  \BibitemOpen
  \bibfield  {author} {\bibinfo {author} {\bibfnamefont {L.}~\bibnamefont {Johns}}\ and\ \bibinfo {author} {\bibfnamefont {S.}~\bibnamefont {Rodriguez}},\ }\bibfield  {title} {\bibinfo {title} {Collisional flavor pendula and neutrino quantum thermodynamics},\ }\href@noop {} {\  (\bibinfo {year} {2023})},\ \Eprint {https://arxiv.org/abs/2312.10340} {arXiv:2312.10340 [hep-ph]} \BibitemShut {NoStop}%
\bibitem [{\citenamefont {Johns}(2025)}]{johns2024}%
  \BibitemOpen
  \bibfield  {author} {\bibinfo {author} {\bibfnamefont {L.}~\bibnamefont {Johns}},\ }\bibfield  {title} {\bibinfo {title} {Subgrid modeling of neutrino oscillations in astrophysics},\ }\href {https://doi.org/10.1103/3fr2-qttd} {\bibfield  {journal} {\bibinfo  {journal} {Phys. Rev. D}\ }\textbf {\bibinfo {volume} {112}},\ \bibinfo {pages} {043024} (\bibinfo {year} {2025})}\BibitemShut {NoStop}%
\bibitem [{\citenamefont {Johns}\ and\ \citenamefont {Kost}(2025)}]{johns2025local}%
  \BibitemOpen
  \bibfield  {author} {\bibinfo {author} {\bibfnamefont {L.}~\bibnamefont {Johns}}\ and\ \bibinfo {author} {\bibfnamefont {A.}~\bibnamefont {Kost}},\ }\bibfield  {title} {\bibinfo {title} {Local-equilibrium theory of neutrino oscillations},\ }\href@noop {} {\bibfield  {journal} {\bibinfo  {journal} {arXiv preprint arXiv:2506.03271}\ } (\bibinfo {year} {2025})}\BibitemShut {NoStop}%
\bibitem [{\citenamefont {Johns}\ \emph {et~al.}(2020)\citenamefont {Johns}, \citenamefont {Nagakura}, \citenamefont {Fuller},\ and\ \citenamefont {Burrows}}]{johns2020}%
  \BibitemOpen
  \bibfield  {author} {\bibinfo {author} {\bibfnamefont {L.}~\bibnamefont {Johns}}, \bibinfo {author} {\bibfnamefont {H.}~\bibnamefont {Nagakura}}, \bibinfo {author} {\bibfnamefont {G.~M.}\ \bibnamefont {Fuller}},\ and\ \bibinfo {author} {\bibfnamefont {A.}~\bibnamefont {Burrows}},\ }\bibfield  {title} {\bibinfo {title} {Neutrino oscillations in supernovae: Angular moments and fast instabilities},\ }\href {https://doi.org/10.1103/PhysRevD.101.043009} {\bibfield  {journal} {\bibinfo  {journal} {Phys. Rev. D}\ }\textbf {\bibinfo {volume} {101}},\ \bibinfo {pages} {043009} (\bibinfo {year} {2020})}\BibitemShut {NoStop}%
\bibitem [{\citenamefont {Zaizen}\ and\ \citenamefont {Nagakura}(2023)}]{zaizen2023}%
  \BibitemOpen
  \bibfield  {author} {\bibinfo {author} {\bibfnamefont {M.}~\bibnamefont {Zaizen}}\ and\ \bibinfo {author} {\bibfnamefont {H.}~\bibnamefont {Nagakura}},\ }\bibfield  {title} {\bibinfo {title} {Characterizing quasisteady states of fast neutrino-flavor conversion by stability and conservation laws},\ }\href {https://doi.org/10.1103/PhysRevD.107.123021} {\bibfield  {journal} {\bibinfo  {journal} {Phys. Rev. D}\ }\textbf {\bibinfo {volume} {107}},\ \bibinfo {pages} {123021} (\bibinfo {year} {2023})}\BibitemShut {NoStop}%
\bibitem [{\citenamefont {Fiorillo}\ \emph {et~al.}(2024)\citenamefont {Fiorillo}, \citenamefont {Raffelt},\ and\ \citenamefont {Sigl}}]{fiorillo2024}%
  \BibitemOpen
  \bibfield  {author} {\bibinfo {author} {\bibfnamefont {D.~F.~G.}\ \bibnamefont {Fiorillo}}, \bibinfo {author} {\bibfnamefont {G.~G.}\ \bibnamefont {Raffelt}},\ and\ \bibinfo {author} {\bibfnamefont {G.}~\bibnamefont {Sigl}},\ }\bibfield  {title} {\bibinfo {title} {Inhomogeneous kinetic equation for mixed neutrinos: Tracing the missing energy},\ }\href {https://doi.org/10.1103/PhysRevLett.133.021002} {\bibfield  {journal} {\bibinfo  {journal} {Phys. Rev. Lett.}\ }\textbf {\bibinfo {volume} {133}},\ \bibinfo {pages} {021002} (\bibinfo {year} {2024})}\BibitemShut {NoStop}%
\bibitem [{\citenamefont {Banerjee}\ \emph {et~al.}(2011)\citenamefont {Banerjee}, \citenamefont {Dighe},\ and\ \citenamefont {Raffelt}}]{banerjee2011}%
  \BibitemOpen
  \bibfield  {author} {\bibinfo {author} {\bibfnamefont {A.}~\bibnamefont {Banerjee}}, \bibinfo {author} {\bibfnamefont {A.}~\bibnamefont {Dighe}},\ and\ \bibinfo {author} {\bibfnamefont {G.}~\bibnamefont {Raffelt}},\ }\bibfield  {title} {\bibinfo {title} {Linearized flavor-stability analysis of dense neutrino streams},\ }\href {https://doi.org/10.1103/PhysRevD.84.053013} {\bibfield  {journal} {\bibinfo  {journal} {Phys. Rev. D}\ }\textbf {\bibinfo {volume} {84}},\ \bibinfo {pages} {053013} (\bibinfo {year} {2011})}\BibitemShut {NoStop}%
\bibitem [{\citenamefont {Chakraborty}\ \emph {et~al.}(2016{\natexlab{b}})\citenamefont {Chakraborty}, \citenamefont {Hansen}, \citenamefont {Izaguirre},\ and\ \citenamefont {Raffelt}}]{chakraborty2016}%
  \BibitemOpen
  \bibfield  {author} {\bibinfo {author} {\bibfnamefont {S.}~\bibnamefont {Chakraborty}}, \bibinfo {author} {\bibfnamefont {R.}~\bibnamefont {Hansen}}, \bibinfo {author} {\bibfnamefont {I.}~\bibnamefont {Izaguirre}},\ and\ \bibinfo {author} {\bibfnamefont {G.}~\bibnamefont {Raffelt}},\ }\bibfield  {title} {\bibinfo {title} {Collective neutrino flavor conversion: Recent developments},\ }\href {https://doi.org/http://dx.doi.org/10.1016/j.nuclphysb.2016.02.012} {\bibfield  {journal} {\bibinfo  {journal} {Nucl. Phys.}\ }\textbf {\bibinfo {volume} {B908}},\ \bibinfo {pages} {366 } (\bibinfo {year} {2016}{\natexlab{b}})}\BibitemShut {NoStop}%
\bibitem [{\citenamefont {Lichtenberg}\ and\ \citenamefont {Lieberman}(1992)}]{lichtenberg1992regular}%
  \BibitemOpen
  \bibfield  {author} {\bibinfo {author} {\bibfnamefont {A.~J.}\ \bibnamefont {Lichtenberg}}\ and\ \bibinfo {author} {\bibfnamefont {M.~A.}\ \bibnamefont {Lieberman}},\ }\href@noop {} {\emph {\bibinfo {title} {Regular and Chaotic Dynamics}}},\ \bibinfo {edition} {2nd}\ ed.,\ Vol.~\bibinfo {volume} {38}\ (\bibinfo  {publisher} {Springer Science \& Business Media},\ \bibinfo {year} {1992})\BibitemShut {NoStop}%
\bibitem [{\citenamefont {Fiorillo}\ and\ \citenamefont {Raffelt}(2025{\natexlab{b}})}]{fiorillo2025collective}%
  \BibitemOpen
  \bibfield  {author} {\bibinfo {author} {\bibfnamefont {D.~F.~G.}\ \bibnamefont {Fiorillo}}\ and\ \bibinfo {author} {\bibfnamefont {G.~G.}\ \bibnamefont {Raffelt}},\ }\bibfield  {title} {\bibinfo {title} {Collective flavor conversions are interactions of neutrinos with quantized flavor waves},\ }\href {https://doi.org/10.1103/PhysRevLett.134.211003} {\bibfield  {journal} {\bibinfo  {journal} {Phys. Rev. Lett.}\ }\textbf {\bibinfo {volume} {134}},\ \bibinfo {pages} {211003} (\bibinfo {year} {2025}{\natexlab{b}})}\BibitemShut {NoStop}%
\end{thebibliography}%

\end{document}